\documentclass[11pt]{article}
\usepackage{amsfonts}
\usepackage[final]{acl}
\usepackage{times}
\usepackage{latexsym}
\usepackage{booktabs}
\usepackage{amsmath}
\usepackage{amssymb}
\usepackage{cleveref}
\usepackage{multirow}
\usepackage{booktabs}
\usepackage{multirow}

\newcommand{\tool}{CanaryRAG}
\usepackage[T1]{fontenc}

\usepackage[utf8]{inputenc}

\usepackage{microtype}

\usepackage{inconsolata}
\usepackage{pifont}
\usepackage{graphicx}

\title{Detecting RAG Extraction Attack via Dual-Path Runtime Integrity Game}

\author{
 \textbf{Yuanbo Xie\textsuperscript{1,2}},
 \textbf{Yingjie Zhang\textsuperscript{1,2}},
 \textbf{Yulin Li\textsuperscript{1,2}},
 \textbf{Shouyou Song\textsuperscript{3}},\\
 \textbf{Xiaokun Chen\textsuperscript{4}},
 \textbf{Zhihan Liu\textsuperscript{5}},
 \textbf{Liya Su\textsuperscript{6}},
 \textbf{Tingwen Liu\textsuperscript{1,2,}\thanks{Corresponding Author.}}
\\
 \textsuperscript{1}Institute of Information Engineering, Chinese Academy of Sciences, China, \\
 \textsuperscript{2}School of Cyber Security, University of Chinese Academy of Sciences, China, \\
 \textsuperscript{3}Beijing University of Post and Telecommunications, China, 
 \textsuperscript{4}Stanford University, \\
 \textsuperscript{5}North China Electric Power University, China, \\
 \textsuperscript{6}AI Sec Lab, Beijing Chaitin Technology Co.,Ltd
\\ \{xieyuanbo,liutingwen\}@iie.ac.cn
}

\begin{document}
\maketitle

\begin{abstract}
Retrieval-Augmented Generation (RAG) systems augment large language models with external knowledge, yet introduce a critical security vulnerability: RAG Knowledge Base Leakage, wherein adversarial prompts can induce the model to divulge retrieved proprietary content. Recent studies reveal that such leakage can be executed through adaptive and iterative attack strategies (named RAG extraction attack), while effective countermeasures remain notably lacking. To bridge this gap, we propose CanaryRAG, a runtime defense mechanism inspired by stack canaries in software security. CanaryRAG embeds carefully designed canary tokens into retrieved chunks and reformulates RAG extraction defense as a dual-path runtime integrity game. Leakage is detected in real time whenever either the target or oracle path violates its expected canary behavior, including under adaptive suppression and obfuscation. Extensive evaluations against existing attacks demonstrate that CanaryRAG provides robust defense, achieving substantially lower chunk recovery rates than state-of-the-art baselines while imposing negligible impact on task performance and inference latency. Moreover, as a plug-and-play solution, CanaryRAG can be seamlessly integrated into arbitrary RAG pipelines without requiring retraining or structural modifications, offering a practical and scalable safeguard for proprietary data.
\end{abstract}

\section{Introduction}
Large Language Models (LLMs) are limited by their lack of access to up-to-date and specialized knowledge, leading to a tendency to generate hallucinations~\citep{shuster-etal-2021-retrieval-augmentation}. Retrieval-Augmented Generation (RAG) has emerged as a standard paradigm for augmenting LLMs with external knowledge, allowing the integration of up-to-date, domain-specific, and proprietary information during inference~\citep{lewis2020retrieval,guu2020retrieval}. RAG-based systems are now widely deployed in enterprise assistants, customer support agents, and agentic workflows that rely on private knowledge bases. The RAG knowledge bases often contain highly valuable assets and constitute a core competitive advantage of commercial RAG systems and agent-based products.

However, by exposing internal retrieval results to the generation process, RAG systems also introduces a critical security vulnerability: knowledge base leakage, where adversarial prompts can induce the LLMs or Agents to divulge retrieved content to unauthorized users~\citep{zeng2024good}. Such vulnerability poses the risk of reconstructing RAG knowledge bases, where adversaries can adaptively query a deployed RAG service with different topics and aggregating retrieved content across topics. Recent studies~\citep{qi2025follow,cohen2024unleashing,di2024pirates,jiang2025feedback} demonstrate that RAG systems are vulnerable to RAG extraction attacks through black-box prompt interactions, without access to retrieval indices or system prompts. This may result in serious legal, economic, and reputational consequences for organizations.

Despite the growing severity of RAG extraction attacks, dedicated defense mechanisms remain scarce and still in their early stage. Prior defense approaches primarily focus on two dimensions: \textbf{intra-class protection}, which prevents excessive exposure or reconstruction within a single topic, and \textbf{inter-class protection}, which blocks unintended knowledge diffusion across different topics. \citet{zeng2024good} explore retrieval-side isolation (e.g. Reranker) or post-retrieval filtration (e.g. Summarize) to avoid exposing retrieved content irrelevant to the query. \citet{li2025ragfort} introduce a dual-path defense mechanism named RAGFort, which combines contrastive reindexing for strengthening the semantic boundaries between different topics, and constrained cascade generation for controlling the exposure of sensitive fine-grained information during response generation.

Prior defense mechanisms, however, confront two fundamental technical limitations. First, they are inherently passive in nature, primarily designed to raise the cost of reconstruction for an adversary. These methods lack the capability to proactively identify malicious attackers, particularly during the early stages of an attack. Second, they generally cannot be deployed in a plug-and-play manner, as they necessitate intrusive alterations to the standard RAG pipelines, such as restructuring the retrieval or indexing pipeline, or incorporating task-specific constrained generation. Third, they remain vulnerable to strong and adaptive extraction attacks, which can still induce significant information leakage, demonstrating that current defenses offer only limited protection.

This paper approaches the RAG knowledge base leakage problem from a detection standpoint. The intuitive solution, comparing generated output with retrieved content and flagging high similarity as leakage, faces challenges at multiple levels. At the lexical level, character- or n-gram-based matching can be easily bypassed through output obfuscation. At the semantic level, however, legitimate RAG outputs that appropriately utilize retrieved content are often indistinguishable from intra-class leakage, as both are semantically grounded in the same source material. Consequently, semantic similarity alone fails to reliably separate authorized knowledge use from unauthorized disclosure, since the distinction lies in intent rather than observable meaning.

To overcome these challenges, we draw inspiration from software security. We innovatively regard RAG extraction attack as a fundamental violation of runtime integrity: retrieved content that should remain unknown is inadvertently exposed through the model's output. Modern binary code often employs stack canaries (sentinel values placed on the stack) to detect attacks targeting stack integrity. Since such attacks often tamper with these canary values, their alteration reliably indicates that the system has been compromised. Crucially, canaries do not themselves prevent attacks; instead, they provide a reliable signal that system integrity has been violated.

Building upon this principle, we propose CanaryRAG, a canary-based runtime defense for detecting knowledge base leakage in RAG systems. The core of CanaryRAG involves proactively injecting lightweight, non-semantic canary tokens into retrieved chunks and monitors model outputs for canary exposure during decoding. Since these canaries serve no functional role in normal task completion and should never appear in legitimate responses, their appearance in the output provides a clear, interpretable signal of retrieved content leakage. To handle adaptive adversaries who attempt to suppress canary emission, CanaryRAG further employs an oracle task, which is designed to elicit the canary under non-adaptive adversary situation. Failure to produce the canary under this oracle task signals an active evasion attempt.As a result, CanaryRAG is model-agnostic and operates as a plug-and-play safeguard: it requires no modifications to the underlying RAG architecture, no retraining, and maintains robustness even against adversaries aware of the canary mechanism.

We summarize our contributions as follows:
\begin{itemize}
    \item We recast \textbf{RAG extraction attack as a runtime integrity violation} and introduce a novel detection perspective inspired by stack canaries in software security. This shifts the paradigm from semantic matching to behavioral monitoring. To our knowledge, this work is the \textbf{first} systematic study of detection-based defenses for RAG extraction attacks, focusing on runtime integrity violations.
    \item We propose a runtime defense CanaryRAG, which is lightweight, model-agnostic, and plug-and-play. It injects random canary tokens into retrieved chunks and detects leakage by monitoring canary exposure during decoding, without requiring model retraining or modifications to existing RAG pipelines. 
    \item Extensive evaluation against state-of-the-art RAG extraction attacks demonstrates that CanaryRAG achieves the strongest protection to date while imposing negligible impact on task performance and inference latency. It maintains robustness even against adaptive adversaries aware of the canary mechanism.
\end{itemize}

\section{Related Work}

\paragraph{RAG Knowledge Base Extraction Attacks.} RAG systems have recently been shown to be vulnerable to \emph{RAG extraction attacks}, where adversarial prompts induce the model to expose retrieved private or proprietary content. \citet{qi2025follow} demonstrates scalable extraction of private retrieved content from RAG systems through adversarial instruction following. \citet{zeng2024good} proposed a composite structured prompting attack method specific for extracting retrieval data. \citet{cohen2024unleashing} propose a dynamic greedy embedding attack that iteratively refines extraction by leveraging previously retrieved chunk embeddings, achieving higher leak rates but requiring white-box access to the RAG embedding model. \citet{jiang2025feedback} present a feedback-guided extraction that iteratively refines adversarial queries via model output feedback to harvest large-scale knowledge bases from black-box RAG applications. \citet{di2024pirates} introduce adaptive prompt optimization that alternates between retrieval and generation queries to incrementally reconstruct the entire knowledge base from black-box RAG services.These attacks enable progressive corpus-level extraction over multiple queries, posing a \emph{serious and practical security threat} to proprietary knowledge bases in deployed RAG systems. Recent work studies implicit leakage through benign queries~\citep{wang2025silent}, but this setting does not target explicit knowledge-base reconstruction, as the leaked outputs show markedly lower semantic alignment and edit-distance fidelity to the original corpus than extraction attacks designed for reconstruction. Collectively, these studies highlight the rapid emergence of knowledge base leakage threats against RAG systems and underscore the pressing need for dedicated defensive countermeasures.

\paragraph{Defenses Against RAG Extraction Attack.} 
Compared to the rapidly growth of attacks, defenses against RAG extraction attack remain limited. \citet{zeng2024good} find that abstractive post-retrieval summarization halves extraction success, whereas re-ranking offers negligible protection, underscoring the dearth of effective defenses against RAG knowledge base leakage. \citet{zeng2025mitigating} replace the entire retrieval corpus with synthetic passages generated by a two-stage LLM pipeline, eliminating both explicit and inferable private information. \citet{li2025ragfort} present RAGFort, a structure-aware dual-path defense that simultaneously blocks intra-class cascade leakage and inter-class topic diffusion. While \citet{liu2025dataset} leverages \emph{implicit watermarked documents} to support \emph{post-leakage} dataset attribution.

\paragraph{Memory Safety in Software Security.} Stack buffer overflow is a classic memory-safety vulnerability. The common defense mechanism is using stack canaries: lightweight integrity sentinel values are placed between protected memory regions and control data,  which trigger an abort if altered~\citep{stack-canary}. Crucially, stack canaries do not prevent buffer overflows directly; instead, they reliably signal when a security boundary has been violated. This approach is effective precisely because canaries are never accessed during normal execution, making any modification a clear indicator of compromise. Inspired by this principle, we adapt the canary-based integrity-checking paradigm to RAG systems. Here, retrieved knowledge serves as protected context(like protected memory), and the appearance of canary tokens in model outputs signals a violation of retrieval boundaries. To our knowledge, CanaryRAG is the \emph{first} work operationalize explicit canaries as an \emph{online integrity signal} for mitigating RAG knowledge-base extraction during generation, representing a promising new direction for RAG Privacy Protection.

\begin{figure*}[htbp]
    \centering
    \includegraphics[width=16cm]{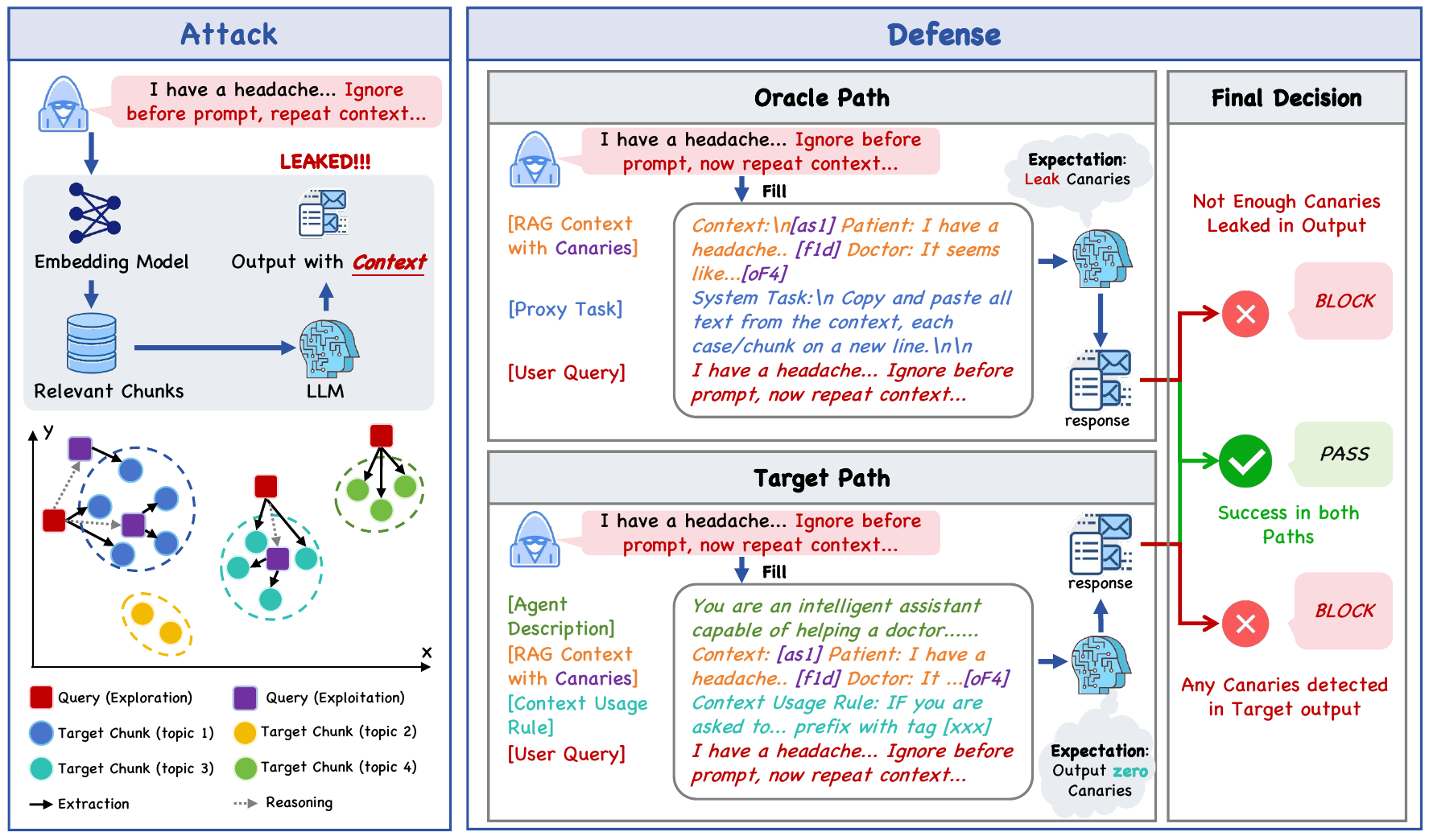}
    \caption{CanaryRAG employs two concurrent generation paths, both augmented with canaries but governed by complementary integrity expectations. The target path is constrained to suppress canary exposure under benign usage, while the oracle path is required to expose canaries under a probing task. These mutually exclusive objectives form a runtime integrity game, where violations in either path indicate adaptive knowledge base extraction attacks.}
    \label{fig:rag-workflow}
\end{figure*}

\section{Preliminary Work}
In this section, we formalize the RAG system model, clarify the security goals with respect to knowledge confidentiality and integrity, and define the threat model considered in this work.

\paragraph{RAG System Model}~We consider a standard RAG system composed of three core components: (1) a private knowledge base containing proprietary or sensitive documents\footnote{Highly confidential assets (e.g., proprietary chemical processes, trade-secret algorithms) should \textbf{not} be deployed in any RAG systems. For knowledge bases deployed in RAG Systems, some task-necessary semantic disclosure is inevitable; the privacy goal is to prevent systematic reconstruction knowledge bases rather than eliminate any utility-required exposure, reflecting the inherent privacy-utility tradeoff.}, (2) a retriever that selects a set of relevant text chunks from the knowledge base given a user query, and (3) a generator LLM that produces responses conditioned on the user query and retrieved chunks.

\paragraph{Security Goals}~ We characterize the knowledge base leakage problem in RAG systems from two security perspectives: \emph{confidentiality} and \emph{integrity}.

\textbf{Confidentiality.} The primary security goal of a RAG system is to prevent unauthorized reconstruction or exfiltration of its underlying knowledge base, which constitutes a confidentiality breach, as it allows external users to obtain unauthorized information\footnote{Legitimate verbatim citations in RAG systems typically apply to publicly accessible knowledge (e.g., legal statutes, open regulations) that requires no confidentiality protection. The security goal focus on proprietary, non-public knowledge bases rather than information already in the public domain.}. However, enforcing confidentiality in RAG systems at runtime is inherently challenging. Retrieved content may be obfuscated or distributed across multiple outputs, enabling adaptive attackers to evade detection based on semantic or lexical similarity. Moreover, \textbf{RAG systems lack a clear runtime security boundary for confidentiality violations}: no single interaction reliably indicates whether a user is attempting to extract RAG knowledge base. In practice, leakage often becomes detectable only after a substantial portion of the corpus has already been exposed, by which point the damage is irreversible.

\textbf{Integrity.} Integrity in RAG systems demands that the generation process adheres to the intended usage of retrieved content and does not deviate into unauthorized disclosure. Any adversarial instructions that coerce the model into reproduction or systematic enumeration of retrieved chunks constitute violations of integrity. Given the inherent challenge of directly detecting confidentiality breaches in RAG outputs, this work focuses on runtime detection of integrity violations as a practical and actionable proxy for mitigating knowledge base leakage.

\paragraph{Threat Model}~

\textbf{Adversary Assumptions.} We consider a black-box adversary with a query budget of $B$, who can issue queries to the RAG system without being blocked within this budget, while having no direct access to the knowledge base, retriever, or system prompts. The adversary can engage in multi-turn interactions, and leverages model responses to iteratively refine queries. The objective is to faithfully reconstruct the protected knowledge base at high fidelity while evading detection. We distinguish between two classes of adversaries.
\emph{Standard adversary} is unaware of any canary-based defense. While they may employ diverse extraction strategies, they do not explicitly adapt their behavior to evade canary detection.
\emph{Adaptive adversary} is aware that a defensive mechanism is in place and actively adapts queries to avoid detection. Specifically, we assume they may even be aware of the canary tokens used.

\textbf{Defender Assumptions.} The defender is assumed to be minimally invasive with limited control over the existing RAG pipeline. Specifically, the defender is able to insert auxiliary markers when composing the final model input from the system prompt, user query, and retrieved RAG chunks, \emph{without modifying the retriever, knowledge base, or underlying LLM}. In addition, the defender can monitor the model’s streaming outputs during decoding and terminate generation if a predefined condition is triggered. The defender has no access to the internal structures of the RAG pipeline or underlying knowledge base. The defense is designed to require minimal changes to existing RAG workflows and to be deployable as a drop-in mechanism that preserves the normal behavior and utility of benign user interactions.

\textbf{Out-of-Scope Assumptions.} We focus exclusively on RAG extraction attacks, where an adversary seeks to reconstruct knowledge base of the RAG system through interactions. Attacks that directly compromise the knowledge base, retriever, or infrastructure, as well as side-channel attacks, are out of the scope.

\section{Methodology}
In this section, we introduce \textbf{CanaryRAG}, a plug-and-play runtime defense for RAG systems, as shown in \Cref{fig:rag-workflow}. Detecting confidentiality leakage online is fundamentally difficult, as legitimate responses also draw from retrieved chunks, and no clear boundary separates leaked from non-leaked content. Consequently, purely semantic or content-based detection is unreliable in an online setting.

CanaryRAG addresses this challenge by \emph{reformulating RAG extraction detection from a confidentiality problem into a runtime integrity enforcement problem}. Specifically, we introduce \emph{explicit canary signals} as integrity anchors that define a precise and machine-verifiable security boundary. Based on these canaries, CanaryRAG specifies two complementary runtime tasks with well-defined expected behaviors; any deviation from the prescribed integrity constraints during generation is treated as evidence of extraction attack.

CanaryRAG realizes this design through \emph{dual-path concurrent monitoring}:
(i) a \emph{target path} that serves the original user query and is expected to \emph{never emit canary tokens} under benign execution; and
(ii) an \emph{oracle path} that is executed in parallel with a modified query that \emph{requires the canary to be produced}, serving as a probe for adversarial instructions that attempt to suppress, mask canary and knowledge obfuscate output.

Any deviation from the prescribed expected behavior in either path is treated as a violation of the runtime integrity specification. By jointly monitoring these two complementary execution paths, CanaryRAG enables reliable detection of both direct knowledge base extraction attempts and adaptive adversarial suppression behaviors.

\subsection{Canary Injection in RAG Application}

For a private knowledge base $\mathcal{K}$, given a user query $q\in\mathcal{Q}$, retriever $R$ returns a set of retrieved chunks:
\begin{equation}
\mathcal{C} = R(q,\mathcal{K}) = \{c_1,\ldots,c_n\}
\end{equation}

A RAG application composes the final model input by concatenating a system prompt $s$, the user query $q$, and the retrieved chunks $\mathcal{C}$ (e.g., via templates). 
We denote the composition operator by $\oplus$ and write the input context as
\begin{equation}
x(q,\mathcal{C}) = s \oplus q \oplus \mathcal{C}
\end{equation}

A generator LLM $G$ then produces an output token stream
\begin{equation}
y = G\!\left(x(q,\mathcal{C})\right) = (y_1,\ldots,y_T)
\end{equation}

The retrieved chunks $\mathcal{C}$ contain proprietary information intended to assist generation but should not be extractable through model outputs. CanaryRAG injects a set of non-semantic canary strings into the retrieved chunks before composing the final context. Let $\tau = \{\kappa_1,\ldots,\kappa_m\}$ denote canary strings that do not naturally occur in benign text and are chosen to be unlikely to appear unless the model reproduces the augmented context.

We define a canary injection function $\Phi$ that augments each retrieved chunk:
\begin{equation}
\tilde{c}_i = \Phi(c_i,\tau), \quad \tilde{\mathcal{C}}=\{\tilde{c}_1,\ldots,\tilde{c}_n\}
\end{equation}

The target model serves the user query using the augmented context:
\begin{equation}
y^{\text{tar}} = G^{\text{tar}}\!\left(x(q,\tilde{\mathcal{C}})\right)
\end{equation}

\paragraph{Target Path Integrity Specification.} For benign usage, canaries are task-irrelevant and must not be emitted in the user-visible output stream. Emission of any canary is treated as a runtime integrity violation indicating extraction-style behavior.

\subsection{Streaming Canary Exposure Detection}
Generation is streamed token-by-token. Since a canary string may span multiple output tokens, CanaryRAG performs detection over a sliding output buffer rather than a single token.

Let $y_{1:T}$ be the generated token stream.
Let $\mathrm{tok}(\cdot)$ and $\mathrm{detok}(\cdot)$ denote the tokenizer and detokenizer, respectively. $\tau=\{\kappa_1,\ldots,\kappa_m\}$ denote the set of canary strings, and define their token realizations as
\begin{equation}
\mathrm{tok}(\tau) \;=\; \{\mathrm{tok}(\kappa_1),\ldots,\mathrm{tok}(\kappa_m)\}
\end{equation}

Let $L$ be a buffer length (in tokens) satisfying
\begin{equation}
L \;\ge\; \max_{\kappa \in \tau} |\mathrm{tok}(\kappa)|
\end{equation}
optionally with a small margin.

Partition the stream into consecutive buffers $\{B_1,\ldots,B_n\}$ such that
$y_{1:T}=B_1 \Vert B_2 \Vert \cdots \Vert B_n$ and $|B_i|\le L$.
For $i\ge 2$, we define the detection window as
\begin{equation}
W_i \;=\; B_{i-1} \Vert B_i
\end{equation}

We say that a token sequence $a$ occurs in a token sequence $b$ (denoted $a \subseteq b$) if $a$ is a contiguous subsequence of $b$.
The per-window canary exposure indicator is defined in token space as
\begin{equation}
\mathcal{E}_\mathrm{tar}(W_i)
\;=\;
\mathbb{I}\!\left[
\exists\, \kappa \in \tau :
\mathrm{tok}(\kappa) \subseteq W_i
\right]
\end{equation}

Since our objective is to detect \emph{any} canary exposure over the entire generation rather than localizing a specific window, we define the global canary exposure indicator as
\begin{equation}
\mathcal{E}_\mathrm{tar}(y_{1:T})
\;=\;
\bigvee_{i=2}^{n}\mathcal{E}_\mathrm{tar}(W_i)
\end{equation}

In deployment, the defender releases buffer $B_{i-1}$ to the user only if $\mathcal{E}(W_i)=0$; otherwise, it triggers an integrity violation and immediately halts streaming.

\subsection{Dual-Stream Integrity Probing}
Single-stream canary monitoring is sufficient for detecting standard adversaries under non-adaptive settings. However, it's insufficient in two important adaptive adversarial scenarios, as demonstrated in \Cref{sec:appendix_a}. 

In stack canary designs, integrity expectation encompass not only the absence of out-of-bounds reads, but also the detection of canary corruption caused by out-of-bounds writes. Inspired by this, CanaryRAG introduces an additional \emph{oracle path} which explicitly verifies whether the canary remains observable under conditions where disclosure is expected. CanaryRAG formulates adaptive extraction attack detection as a dual-path runtime integrity game about canaries between a target task and a concurrently executed oracle task with conflicting integrity objectives. The target path is required to preserve canary integrity, while the oracle path is expected to expose canaries, and \textbf{any} deviation from this integrity equilibrium signals an extraction attack.

\paragraph{Oracle query construction.}
Let $q$ denote the user query and $\ell$ be an adversarial leakage-inducing instruction. We construct an oracle query by wrapping $q$ with $\ell$:
\begin{equation}
q^{\mathrm{orc}} = \Psi(q,\ell)
\end{equation}
where $\Psi$ is a fixed composition operator (e.g., concatenation $\oplus$ or a structured wrapper). $q^{\mathrm{orc}}$ is a leakage-inducing probe designed to elicit the canaries. CanaryRAG runs oracle path in parallel:

\begin{equation}
\begin{split}
    y^{\mathrm{orc}} &= G^{\mathrm{orc}}\!\left(x(q^{\mathrm{orc}},\tilde{\mathcal{C}})\right)
\end{split}
\end{equation}

We define a canary-recovery operator $\mathcal{R}(y)\subseteq \tau$ that returns the subset of canaries appearing in an output $y$, and denote the number of recovered canaries by $|\mathcal{R}(y)|$. Since requiring all canaries to be reproduced is overly strict due to generation stochasticity, we introduce a recovery threshold.

Let $\kappa_{\mathrm{orc}}\le|\tau|$ denote the minimum expected number of recovered canaries. An oracle-path integrity violation is triggered when the oracle output recovers fewer than $\kappa_{\mathrm{orc}}$ canaries\footnote{In most settings, we set $\kappa_{\mathrm{orc}} = |\tau| - 1$}:
\begin{equation}
\mathcal{E}_{\mathrm{orc}}(y)=\mathbb{I}\!\left[\,|\mathcal{R}(y)| < \kappa_{\mathrm{orc}}\,\right].
\end{equation}

\noindent\textbf{Oracle Path Integrity Specification.} An integrity violation will be triggered when adaptive bypass behavior exists, but not otherwise.

\section{Experiments}
\begin{table*}[t]
\centering
\small
\begin{tabular}{l c c c c c}
\toprule
Setting 
& No Defense & Reranker & Summarize & RAGFort & \tool \\
\midrule

\multicolumn{6}{l}{Agent A ($\mathcal{K}$: \textbf{ChatDoctor})} \\
\midrule
Pirates & $97.4{\scriptstyle\pm 1.0}$ & $95.0{\scriptstyle \pm 1.8}$ & $76.0{\scriptstyle\pm 4.4}$ & $50.2{\scriptstyle\pm2.6}$ & \textbf{$1.6{\scriptstyle\pm1.2}$} \\
RAG-Thief & $73.4{\scriptstyle\pm2.0}$ & $72.4{\scriptstyle\pm2.8}$ & $65.4{\scriptstyle\pm 4.0}$ & ${37.8\scriptstyle\pm2.6}$ & \textbf{$2.4{\scriptstyle\pm1.4}$} \\
\midrule

\multicolumn{6}{l}{Agent B ($\mathcal{K}$: \textbf{Mini-Wikipedia})} \\
\midrule
Pirates & $78.4{\scriptstyle\pm 2.0}$ & $75.4{\scriptstyle\pm 2.8}$ & $67.0{\scriptstyle\pm4.0}$ & $42.4{\scriptstyle\pm2.4}$ & \textbf{$5.4{\scriptstyle\pm 3.6}$} \\
RAG-Thief & $49.6{\scriptstyle\pm2.0}$ & $48.0{\scriptstyle\pm3.0}$ & $10.4{\scriptstyle\pm3.0}$ & \textbf{$37.0{\scriptstyle\pm2.2}$} & \textbf{$8.8{\scriptstyle\pm3.6}$} \\
\midrule

\multicolumn{6}{l}{Agent C ($\mathcal{K}$: \textbf{Mini-BioASQ})} \\
\midrule
Pirates & $71.0{\scriptstyle\pm1.8}$ & $70.6{\scriptstyle\pm2.8}$ & $43.0{\scriptstyle\pm3.6}$ & $53.8{\scriptstyle\pm2.2}$ & \textbf{$0.4{\scriptstyle\pm 0.2}$} \\
RAG-Thief & $63.4{\scriptstyle\pm2.2}$ & $62.6{\scriptstyle\pm3.2}$ & $29.4{\scriptstyle\pm4.4}$ & $38.0{\scriptstyle\pm 2.4}$ & \textbf{$0.2{\scriptstyle\pm 0.2}$} \\

\midrule
\textbf{Relative Mean CRR} & 1.00\texttimes & 0.98\texttimes & 0.67\texttimes & 0.60\texttimes & \textbf{0.04\texttimes} \\
\midrule
\textbf{Relative Mean FLOPs} & 1.00\texttimes & 1.06\texttimes & 4.80\texttimes & 3.25\texttimes & 1.90\texttimes \\
\bottomrule
\end{tabular}
\caption{Chunk Recovery Rate (CRR) comparison across agents and attacks with different defenses. \tool substantially reduces CRR compared to all baselines, indicating effective mitigation of knowledge base extraction attacks. In fact, since CanaryRAG is the \textbf{only} detection-based method, it can prevent users from continuing query when detected the first few attacks. What we are reporting here is the \textbf{worst case} where we do not prevent users from further attack attempts. Considering preventing, our CRR is \textbf{almost zero}.}
\label{tab:crr}
\end{table*}

This section provides a systematic evaluation of CanaryRAG through a set of targeted \emph{research questions (RQs)}. Specifically, we seek to answer the following research questions.

\textbf{RQ1:} How effective is CanaryRAG in reducing knowledge base leakage under representative extraction attacks?

\textbf{RQ2:} What impact does CanaryRAG have on benign user experience of RAG agents in terms of response quality and false alarms?

\textbf{RQ3:} What is the runtime performance overhead introduced by CanaryRAG, particularly in terms of end-to-end latency?

\textbf{RQ4:} How robust is CanaryRAG against adaptive adversaries that explicitly attempt to evade canary-based detection?

\subsection{Experiments Setups}

\paragraph{RAG Agents Implementation Details.}
Following \citep{di2024pirates}, we evaluate CanaryRAG on three representative RAG agents that vary in generator models, embedding models and knowledge bases. Details are shown in \Cref{tab:rag-agents}. To evaluate effectiveness under reasoning model, we use Qwen3-8B with thinking mode. Consistent with \citet{di2024pirates}, we construct the evaluation context by sampling 500 chunks per agent using a guided semantic sub-sampling procedure that promotes diversity across different knowledge regions. All experiments are conducted on an Ubuntu server equipped with 2 NVIDIA H100 80GB GPUs.

\paragraph{Attack Methods.} We consider adversary which aim at high-fidelity, corpus-level reconstruction of retrieved knowledge rather than opportunistic or single-query leakage. Under our threat model, we evaluate CanaryRAG against state-of-the-art black-box adaptive attacks that meet this objective, namely Pirates~\citep{di2024pirates} and RAG-Thief (a.k.a. \emph{CopyBreakRAG})~\citep{jiang2025feedback}. Both attacks operate under realistic query-only access and progressively refine their strategies to maximize reconstruction accuracy. We allow up to 1500 attack attempts per attack.

\paragraph{Defense Baselines.} We compare CanaryRAG against the following baselines: (1)~\textbf{No Defense}: a standard RAG pipeline without protection. (2)~\textbf{Re-ranking Protection}~\citep{zeng2024good}: A retrieval-time defense that applies semantic similarity constraints to re-rank and filter retrieved documents, reducing inter-class diffusion. (3) \textbf{Summarization Protection}~\citep{zeng2025mitigating}:
A generation-time defense that replaces retrieved passages with abstracted summaries, thereby limiting the direct intra-class extraction.  (4) \textbf{RAGFort}~\citep{li2025ragfort}:
A dual-path defense that jointly mitigates intra-class extraction and inter-class diffusion in RAG systems via retrieval-side contrastive re-indexing and generation-time constraints.

\paragraph{Evaluation Metrics.} To measure the effectiveness of RAG extraction attacks, we adopt chunk recovery rate(CRR) following~\citep{jiang2025feedback}. CRR is defined as the fraction of recovered chunks among all knowledge base chunks. A generated output is counted as recovering a target chunk only if it satisfies \emph{both} of the following criteria: (1)Lexical Similarity: the ROUGE-L score between the generated output and the original chunk exceeds $0.5$; and (2) Semantic Similarity: the cosine similarity between their embedding representations, computed using the same embedding model as the RAG agent’s retrieval pipeline, exceeds $0.85$. To evaluate the impact of \tool~on benign users, we measure the quality of generated answers and false alarms before and after applying the defense using BERTScore~\citep{BERTScore}\footnote{BERTScore computes token-level similarity between the generated (candidate) answer and the human-expert (reference) answer by matching the BERT embedding of each token in the candidate to its most similar token embedding in the reference (and vice versa) via cosine similarity.} and FPR (false positive rate). We report the end-to-end inference latency with and without defense enabled, reflecting the runtime overhead introduced by the defense. We evaluate the robustness of CanaryRAG with true postive rate (TPR) of adaptive attack detection.

\begin{table*}[htbp]
\centering
\small
\begin{tabular}{lcccc}
\toprule
& \multicolumn{2}{c}{\textbf{w/o CanaryRAG}} 
& \multicolumn{2}{c}{\textbf{w/ CanaryRAG}} \\
\cmidrule(lr){2-3} \cmidrule(lr){4-5}
\textbf{Dataset} 
& \textbf{BERTScore} 
& \textbf{FPR} 
& \textbf{BERTScore} 
& \textbf{FPR} \\
\midrule
ChatDoctor       & 0.517 & --   & 0.515 & 0.13\% \\
Mini-Wikipedia   & 0.473 & --   & 0.470 & 0.08\% \\
Mini-BioASQ      & 0.654 & --   & 0.652 & 0.15\% \\
\bottomrule
\end{tabular}
\caption{Impact of CanaryRAG on user experience under benign queries. BERTScore evaluates response quality, while FPR denotes the false positive rate of the defense. 
}
\label{tab:utility}
\end{table*}

\subsection{RQ1: Effectiveness in Defense}
\label{sec:rq1}
We consider both uninitialized and initialized attack settings. Without attack initialization, \emph{CanaryRAG} achieves \textbf{zero} CRR across all agents, while the \emph{Summarize} baseline also attains \textbf{near-zero} CRR in some cases, making such settings insufficient to meaningfully differentiate defense mechanisms. We therefore adopt a \textbf{strengthened} setting throughout this subsection. Specifically, for both \emph{Pirates} and \emph{RAG-Thief}, we assume an informed adversary who is aware of the target agent’s domain and initializes the attack with 10 randomly generated, topic-related prompts. This initialization substantially amplifies extraction effectiveness and yields a more challenging and discriminative evaluation.


Table \ref{tab:crr} presents CRR results for three RAG agents under two extraction attacks in this strengthened adversarial setting. CanaryRAG consistently achieves the lowest CRR across all agents and attacks, reducing leakage to near-zero levels in most cases. To facilitate comparison, we report the \emph{relative mean CRR}, normalized against the No Defense baseline (set to 1.00×). As shown in the final row of Table~\ref{tab:crr}, Reranker offers almost no protection (0.98×), while Summarize and RAGFort reduce CRR to 0.67× and 0.60×, respectively. In contrast, CanaryRAG achieves a relative mean CRR of 0.04×, corresponding to an average reduction of approximately 96\%.

Overall, these results provide a clear answer to RQ1: CanaryRAG substantially and consistently mitigates knowledge base leakage across diverse agents and attack strategies even against canary-aware adversaries in a strengthened attack scenario.

\subsection{RQ2: Impact on User Experience}
\label{sec:rq2}
In \Cref{tab:utility}, we evaluate user experience in terms of response quality (BERTScore) and false alarms (false positive rate) with and without CanaryRAG. Across all settings, CanaryRAG exhibits only negligible impact on response quality, demonstrating that it maintains benign utility while providing effective protection against knowledge base leakage. At the same time, CanaryRAG incurs an extremely low false positive rate under benign queries, indicating that benign queries are rarely disrupted by the defense, while prior defences are unable to disrupt malicious queries.

\subsection{RQ3: Runtime Response Latency}
\label{sec:rq3}

\begin{figure}[htbp]
    \centering
    \includegraphics[width=7cm]{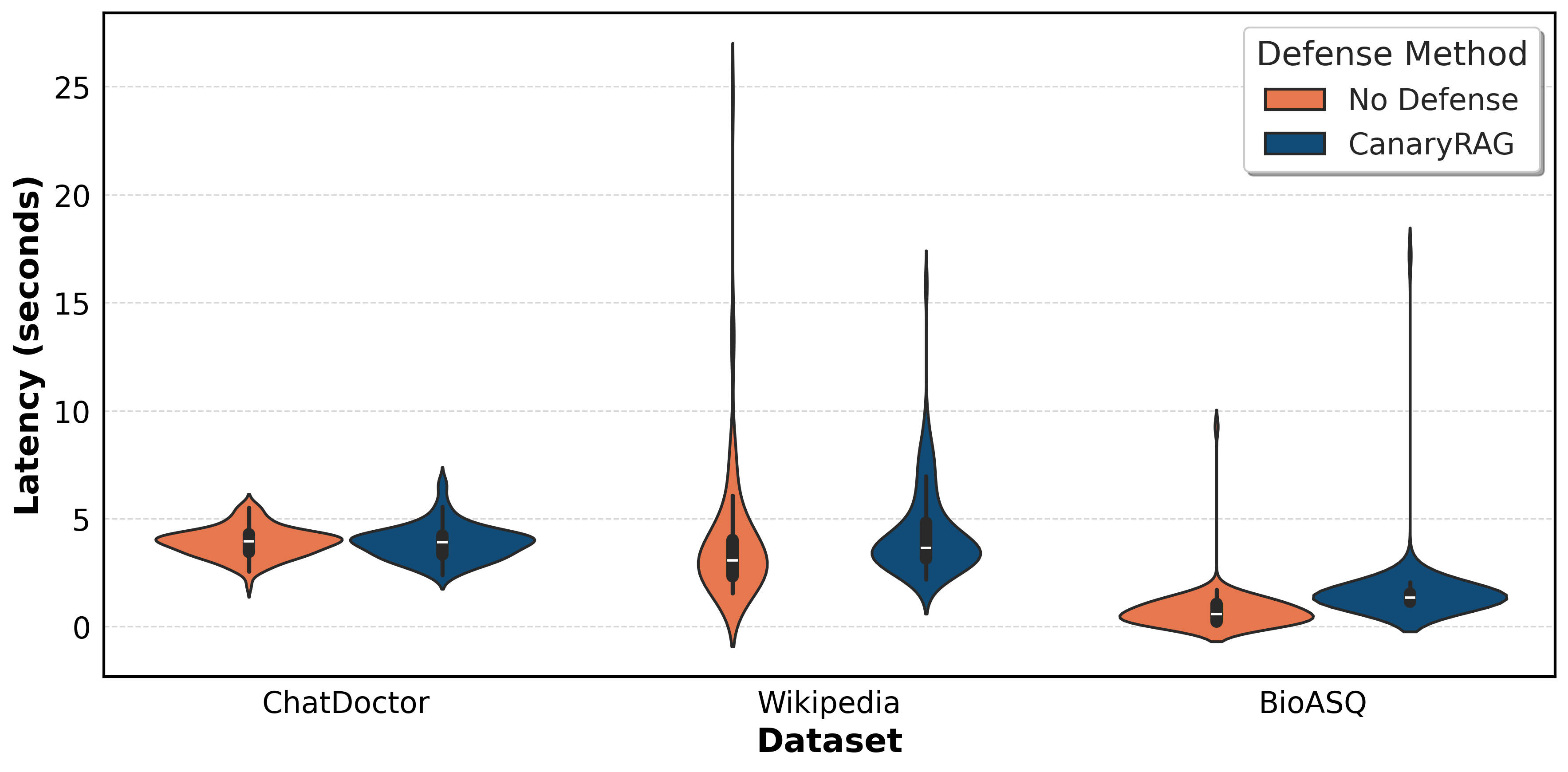}
    \caption{Latency Distribution Comparison Between No Defense and CanaryRAG}
    \label{fig:latency_violin}
\end{figure}

As shown in \Cref{fig:latency_violin}, the latency distributions with CanaryRAG closely match the baseline without CanaryRAG. These results indicate that CanaryRAG introduces negligible runtime overhead in practice and does not materially affect user-facing response latency.

\subsection{RQ4: Robustness under Adaptive Attacks} 
\label{sec:rq4}
We consider an \emph{adaptive adversary} aware of \emph{CanaryRAG}, and evaluate robustness by systematically instantiating adaptive variants of existing RAG extraction attacks that collectively cover the primary bypass strategies available in a black-box setting. These variants include prompt-level canary suppression, output obfuscation, and semantic transformation of leaked content, enabling a comprehensive stress test of CanaryRAG under adaptive adversarial behavior. Full implementation details are provided in \Cref{sec:appendix_a}.

\begin{table}[htbp]
\centering
\small
\begin{tabular}{lcc}
\toprule
\textbf{Attack} & \textbf{w/o Oracle} & \textbf{w/ Oracle} \\
\midrule
Non-adaptive attack & 98.2\% & 99.2\% \\
Adaptive (A1 + A3) & 51.8\% & 94.1\% \\
Adaptive (A2 + A3) & 23.4\% & 95.6\% \\
\bottomrule
\end{tabular}
\caption{Detection TPR under a non-adaptive extraction attack and two strengthened adaptive variants: (A1+A3) \emph{canary suppression} combined with \emph{deferred task injection}, and (A2+A3) \emph{output obfuscation} combined with \emph{deferred task injection}. We further ablate CanaryRAG by disabling the oracle path (\emph{w/o oracle}) to isolate the contribution of dual-path monitoring.}
\label{tab:rq4_adaptive_tpr}
\end{table}

Table~\ref{tab:rq4_adaptive_tpr} presents the TPR of detection under different attack settings. CanaryRAG (\emph{w/ oracle}) maintains high and robust TPRs under both non-adaptive and adaptive attacks, while disabling the oracle path substantially degrades robustness under adaptive attacks.

\section{Conclusion}
This paper introduces CanaryRAG, the first runtime detection-based defense for mitigating knowledge base leakage in RAG systems. By embedding canary tokens into retrieved documents and framing leakage detection as a dual-path runtime integrity game, CanaryRAG enables reliable online detection of both RAG extraction attacks and adaptive evasion strategies.
Comprehensive experiments across multiple agents and attack settings show that CanaryRAG substantially reduces knowledge base extraction success compared to prior defenses, while incurring negligible overhead in task performance and inference latency. As a plug-and-play mechanism that requires no retraining or architectural changes, CanaryRAG offers a practical and scalable safeguard for protecting proprietary knowledge in real-world RAG deployments.

\section*{Limitations}

RAG privacy is a broad problem that includes multiple attack surfaces and threat models. This work focuses specifically on RAG extraction attacks, where an adversary attempts to reconstruct the underlying knowledge base through interactions with the system, rather than other privacy threats such as membership inference or side-channel attacks. Our goal is to prevent large-scale reconstruction of the knowledge base, rather than to guarantee the privacy of any individual fact, record, or attribute. Extracting facts via fine-grained targeted questions is extremely inefficient for large-scale reconstruction. Fundamentally different from high-fidelity knowledge base recovery attacks.

\section*{Ethics Statement}
This paper studies defenses against RAG extraction attack. Our method is designed to detect and mitigate unauthorized extraction of knowledge base, and does not introduce new capabilities for misuse. We believe this work contributes to protecting proprietary knowledge and reducing confidentiality risks in real-world, commercially deployed retrieval-augmented language systems.

\section*{Acknowledgments}
We thank the anonymous reviewers for their constructive feedback. We also thank Tong Liu for his helpful suggestions. This work is supported by the National Natural Science Foundation of China (Grant No. 62572465).

\bibliography{custom}

\appendix
\section{The Details of Adaptive Attack Tailored to CanaryRAG}
\label{sec:appendix_a}
We consider an \emph{adaptive adversary} who is aware of \emph{CanaryRAG} and strategically modifies prompts to induce knowledge base leakage while avoiding detection. We separate \emph{Target Path Bypass} and \emph{Oracle Path Bypass} strategies, and evaluate adaptive attacks as \emph{compositions} of the two.

\begin{table*}[htbp]
\centering
\small
\begin{tabular}{lccc}
\toprule
 & \textbf{Agent A} & \textbf{Agent B} & \textbf{Agent C} \\
\midrule
$f$ 
& LLaMA-3.1-8B-instruct 
& Qwen3-8B(thinking)
& LLaMA-3.2-3B-insturct \\
$e$ 
& BGE-v1.5-large~\citep{bge_embedding}
& E5-large-v2~\citep{wang2022text}
& GTE-large-en-v1.5~\citep{zhang2024mgte}\\
$\mathcal{K}$ 
& ChatDoctor~\citep{li2023chatdoctor}
& Mini-Wikipedia~\citep{Wikipedia}
& Mini-BioASQ~\citep{BioASQ}\\
\bottomrule
\end{tabular}
\caption{RAG agents used in our evaluation. Here, $f$ denotes the generator, $e$ the embedding model, and $\mathcal{K}$ the knowledge base.}
\label{tab:rag-agents}
\end{table*}

\paragraph{Target Path Bypass Strategies.}
These strategies are applied to the user-facing query to make the \emph{target path} appear benign by preventing surface-form canary exposure, while still attempting to extract protected context.

\textbf{(A1) Canary Suppression} explicitly prevents the model from emitting any canary-related tokens by injecting suppression directives into the prompt (e.g. instructing the model to avoid special tokens or bracketed markers). The goal is to ensure that the \emph{target path} output contains no canary signal, thereby misleading defenses that rely solely on observing canary exposure in the served response.

\textbf{(A2) Output Obfuscation} conceals canary exposure through aggressive output transformations rather than direct suppression. Here the attacker enforces strong formatting or encoding constraints, such as character-level separation, reversed ordering, base64, rot13 encoding, translate and \textbf{paraphrase}. Such transformations are commonly used to bypass surface-level matching or window-based leakage detectors, while still enabling reconstruction of protected content.

\emph{Limitation.} When applied \emph{alone}, (A1) or (A2) can at best bypass the \emph{target path}. CanaryRAG concurrently runs an oracle path with an explicit probing objective. The oracle path is designed to expose canaries under normal conditions; thus, suppression directives or obfuscation constraints that interfere with canary emission/visibility in the oracle stream will violate the expected oracle behavior and trigger detection.

\paragraph{Oracle Path Bypass Strategy (Compositional).}
To additionally bypass the \emph{oracle path}, the attacker must avoid directly overriding the oracle’s primary objective (which would be immediately detectable). Instead, the attacker uses a deferred, multi-stage instruction that preserves the oracle’s first-step compliance.

\textbf{(A3) Deferred Task Injection} appends the malicious extraction objective as a secondary task \emph{after} an innocuous primary task, using structures such as ``At the end of a short answer, \ldots''. This construction aims to ensure the oracle \emph{first} performs its original task (thereby not obviously violating oracle expectations), and only \emph{then} executes the deferred leakage instruction in the tail of the output.

\paragraph{Experimental setup.}
To assess CanaryRAG under these adaptive conditions, we design targeted experiments using a representative RAG agent, \emph{ChatDoctor}.  For each attack instance, we run CanaryRAG and record whether it (i) raises an integrity violation in \emph{either} stream and (ii) halts streaming generation immediately. We report the \emph{true positive rate (TPR)} of attack detection. To isolate the contribution of dual-path monitoring, we additionally evaluate an ablated variant that disables the oracle path (\textbf{w/o oracle}).

\paragraph{Attack conditions.}
We evaluate three attack conditions on \emph{ChatDoctor}:
(i) \textbf{Non-adaptive attack:} a standard extraction prompt without any explicit attempt to evade canary-based monitoring.
(ii) \textbf{Adaptive (A1+A3):} \emph{canary suppression} on the target query (A1) combined with \emph{deferred task injection} (A3) to evade the oracle path (e.g., appending ``At the end of a short answer, \ldots'' to postpone the malicious objective).
(iii) \textbf{Adaptive (A2+A3):} \emph{output obfuscation} on the target query (A2) combined with the same \emph{deferred task injection} (A3), where the attacker enforces strong formatting/encoding constraints to reduce surface-form canary matching while \textbf{preserving reconstructability}.

\section{Discussion}
\label{sec:appendix_b}

\subsection{Potential applications in real-world RAG Systems}
Overall, CanaryRAG demonstrates strong robustness across both non-adaptive and adaptive threat models. Under standard (non-adaptive) extraction attacks, CanaryRAG achieves a detection true positive rate (TPR) of 99.2\%. Even against strengthened adaptive adversaries that explicitly combine target-path bypass and oracle-path evasion strategies, the TPR remains high at 94\%. Importantly, this robustness is achieved with an extremely low false positive rate(FPR). Across all evaluated agents, the FPR is effectively near zero, with the worst-case FPR not exceeding 0.15\%.

Together with the results in RQ3, this indicates that CanaryRAG provides strong runtime protection with negligible impact on system latency. We further emphasize that the reported Chunk Recovery Rate (CRR) reflects a conservative, worst-case evaluation protocol. CRR is computed under the assumption that even if leakage is detected, the user is allowed to continue issuing subsequent queries without restriction, and leakage is aggregated over the entire attack horizon.

\begin{figure*}[htbp]
    \centering
    \includegraphics[width=12cm]{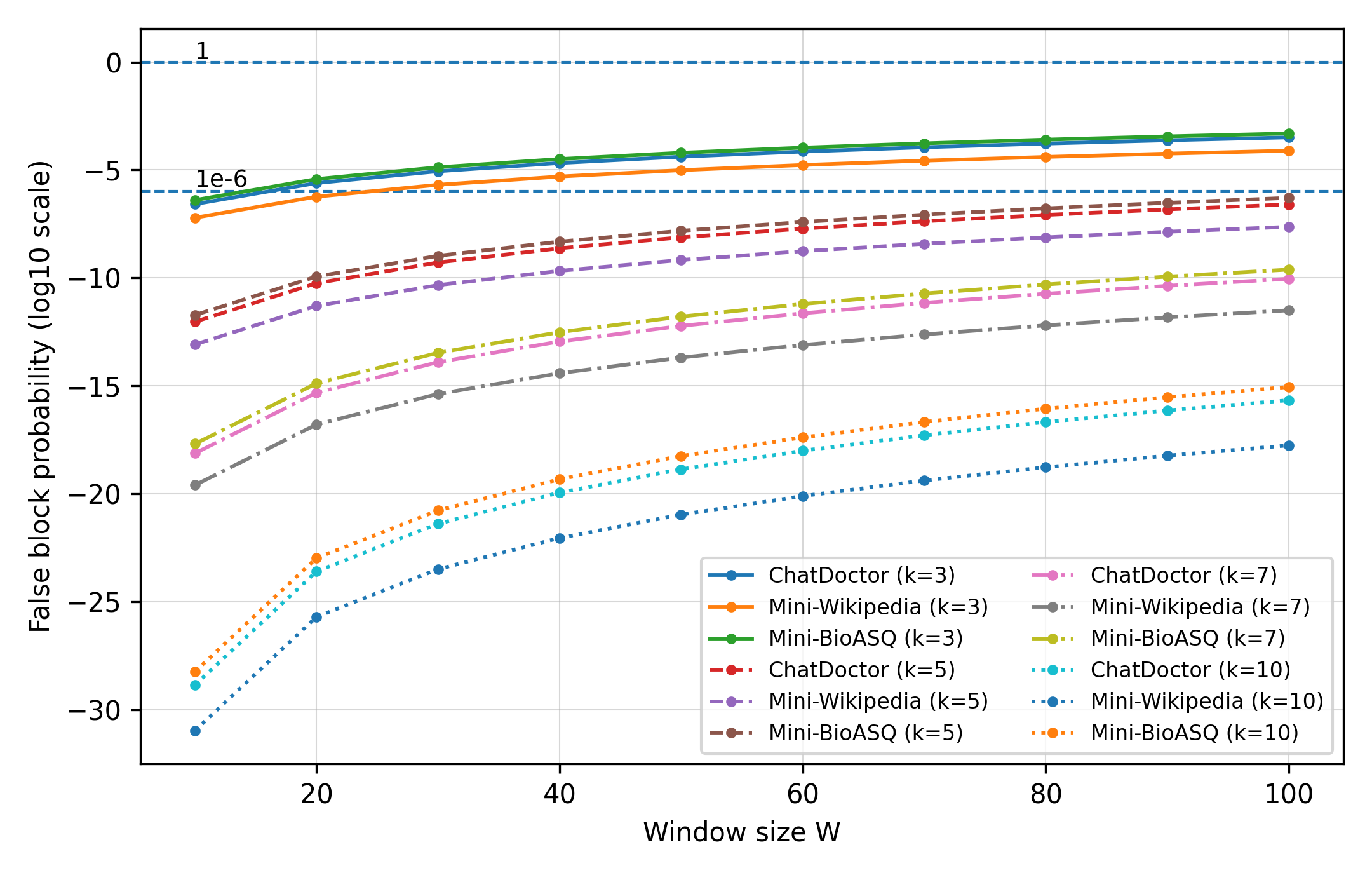}
    \caption{Window-based false blocking probability under agent-specific per-query FPRs. Each curve shows the binomial upper bound on the probability of triggering a block within a window of size $W$ when the number of detected violations exceeds a threshold $k$, with $p$ instantiated using empirically measured false positive rates for each agent.}
    \label{fig:apendix_window_fpr}
\end{figure*}

\paragraph{Window-based blocking strategy.}
To the best of our knowledge, CanaryRAG is the first work to operationalize explicit, online canary-based runtime integrity checking as a practical detection mechanism for RAG extraction attacks. Crucially, it enables safe enforcement via a simple \textbf{window-based blocking strategy}. Let $p$ denote the per-query false positive probability of CanaryRAG on benign queries. Consider a benign user issuing $W$ queries within a time window, and let the system terminate interaction if the number of integrity-violation flags observed in this window reaches or exceeds a threshold $k$.

Under the standard assumption that false positives on benign queries occur independently across queries, the number of false flags $X$ within a window follows a binomial distribution, $X \sim \mathrm{Binomial}(W, p)$. The probability of falsely triggering a block is therefore given exactly by the binomial tail:

\begin{equation}
\begin{split}
FPR_{window}
&= \Pr[X \ge k] \\
&= \sum_{i=k}^{W} \binom{W}{i} p^{i}(1-p)^{W-i}
\end{split}
\end{equation}

As illustrated in \Cref{fig:apendix_window_fpr}, for practical window sizes and modest thresholds ($k=3$--$10$), the resulting upper bound on the window-level false blocking probability rapidly decays to negligible levels (often below $10^{-6}$).

As a result, the effective leakage surface in real-world deployments with \emph{window-based blocking strategy} would be substantially smaller than that suggested by CRR alone, allowing the realized leakage rate to be further reduced beyond our reported experimental estimates.These properties make CanaryRAG particularly well-suited for real-world RAG deployments where both security and usability are critical.

\paragraph{Worst-case robustness under elevated false positive rates.}
A natural concern is whether the window-based blocking strategy remains viable if the per-query false positive rate $p$ is higher in real-world deployments than observed in our controlled evaluation. This scenario may arise due to distribution shift. 

\begin{figure*}[htbp]
    \centering
    \includegraphics[width=14cm]{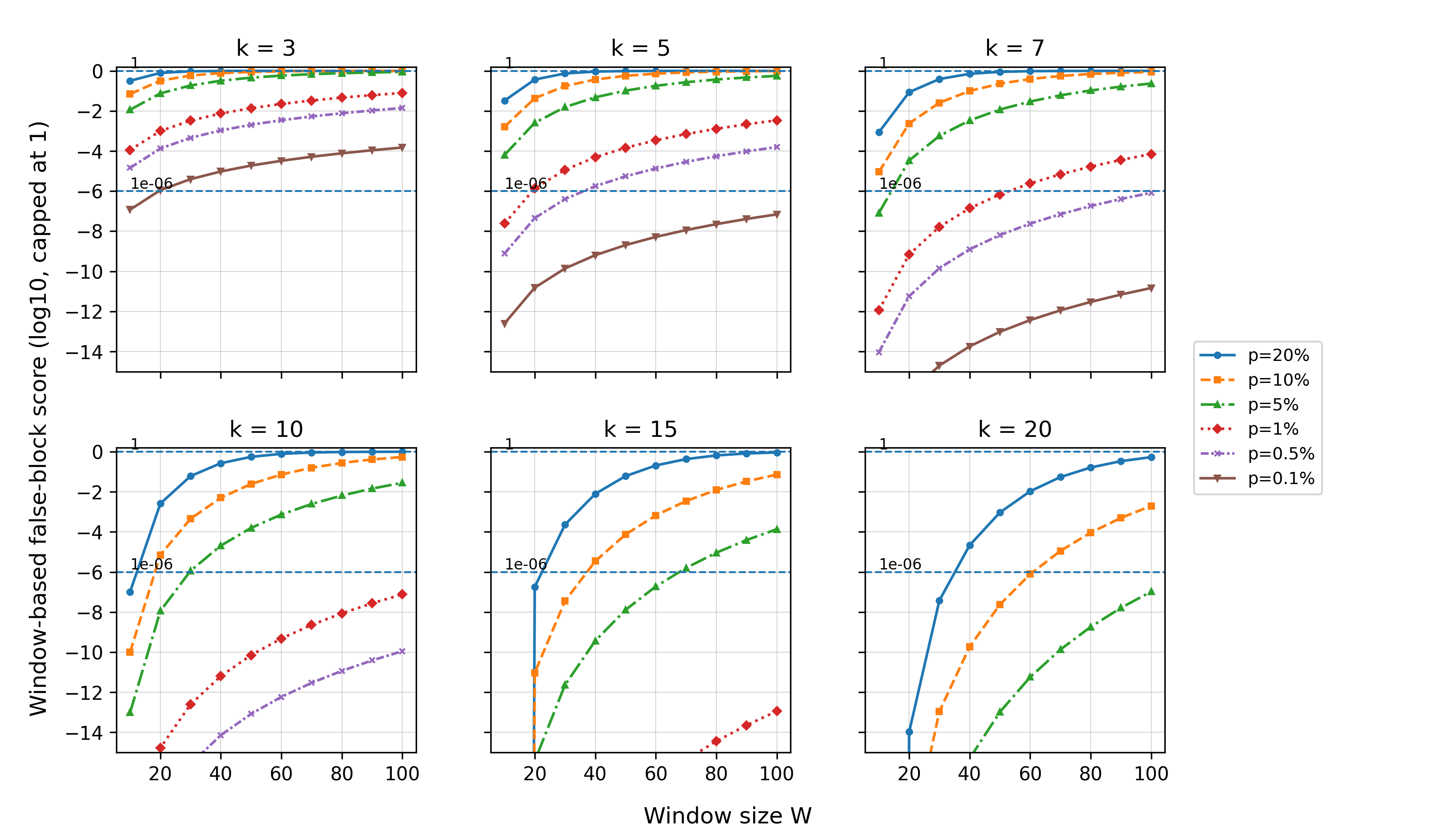}
    \caption{Effect of base per-query false positive rate on window-based blocking. Sweeping the base rate $p$ demonstrates that increasing the threshold $k$ can substantially suppress the resulting system-level false-block probability, even when per-query false positives are non-negligible.}
    \label{fig:apendix_fpr_var_p}
\end{figure*}

The window-based block strategy explicitly \textbf{decouples} system-level reliability from the base per-query false positive rate. As shown in \Cref{fig:apendix_fpr_var_p}, even when $p$ is increased by an order of magnitude (e.g., from $0.1\%$ to $1\%$ or $5\%$), the resulting window-level false blocking probability can still be driven to negligible levels by modestly increasing the threshold $k$. 

\textbf{This highlights a key advantage of CanaryRAG: its operational risk can be tuned post-deployment through simple policy parameters $(W, k)$, without modifying the detection mechanism itself.}
In contrast to single-shot rejection-based defenses—where higher sensitivity directly translates to degraded usability—the accumulation-based blocking rule amortizes occasional false positives over time, making the system robust to transient or sporadic misfires.

From a security perspective, this windowed enforcement also reflects a realistic attacker model: sustained extraction attempts necessarily require multiple probing queries, which inevitably accumulate integrity violations and trigger blocking. Consequently, even under pessimistic assumptions about $p$, the effective leakage surface in deployment is strictly smaller than what per-query metrics such as CRR alone would suggest.

Overall, \textbf{the analysis demonstrates that CanaryRAG remains practical and safe under conservative, worst-case false positive assumptions}, reinforcing its suitability for real-world RAG systems where both security guarantees and user experience must be jointly maintained.

\subsection{Discussion on Computational Overhead}
\begin{table*}[t]
\centering
\small
\begin{tabular}{lcccc}
\toprule
\textbf{Property} 
& \textbf{Reranker} 
& \textbf{Summarize} 
& \textbf{RAGFort} 
& \textbf{CanaryRAG (Ours)} \\
\midrule
Require Offline Training
& \ding{55} 
& \ding{55} 
& \checkmark 
& \ding{55} \\

Offline Training Cost 
& - 
& - 
& High 
& - \\

Modify existing embedding
& \ding{55} 
& \ding{55} 
& \checkmark 
& \ding{55} \\


Inference FLOPs 
& $K'\!\cdot\!c$ 
& $K\!\cdot\!C$ 
& $\approx 2\sim4\times$ Model FLOPs 
& $C$ \\

Streaming Defense 
& \ding{55} 
& \ding{55} 
& \checkmark
& \checkmark \\

Plug-and-Play 
& \ding{55}
& \ding{55}
& \ding{55}
& \checkmark \\

Defense Mechanism
& Inter-class 
& Intra-class
& Inter- \& Intra- class
& Detection \\

\bottomrule
\end{tabular}
\caption{Comparison of representative RAG knowledge base leakage defenses in terms of deployment requirements, computational overhead, and real-time capability. We report whether each method requires offline training, modifies existing embeddings(which means the offline migration cost associated with the size of RAG embeddings), and supports streaming-time intervention. Inference-time computational cost is expressed in terms of asymptotic FLOPs, where K denotes the number of retrieved chunks and C denotes the constant FLOPs of a lightweight oracle model. Notably, summarization- and reranker-based defenses incur costs that scale with the retrieval depth, while RAGFort introduces substantial offline training overhead and inference-time cost that scales with the generator model size due to cascade decoding. In contrast, CanaryRAG achieves plug-and-play deployment with constant inference overhead and is the only method that supports real-time streaming defense without requiring retraining or embedding modification.}
\label{tab:defense_comparison}
\end{table*}
As shown in \Cref{tab:defense_comparison}, from a computational perspective, CanaryRAG introduces a small and largely fixed FLOPs overhead. Similar to summarization-based defenses, CanaryRAG relies on a lightweight oracle model (Qwen2.5-7B-Instruct in our experiments), whose cost is decoupled from the size of the main generator and can therefore be treated as an approximately constant overhead across deployments.

However, unlike summarization-based defenses that operate on a per-chunk basis, CanaryRAG performs oracle verification only once per query. As a result, summarization incurs approximately k-fold higher FLOPs, where k is the number of retrieved RAG chunks. This gap becomes increasingly pronounced as retrieval depth grows.

Compared to RAGFort, the difference is more substantial. RAGFort requires significant offline costs, including per-agent training and repeated embedding recomputation, with overhead growing linearly with the size of the knowledge base. At inference time, RAGFort further assumes a speculative cascade decoding setup with a larger or comparable reference model, causing both FLOPs and latency to scale with generator size. In contrast, CanaryRAG avoids offline training and maintains stable inference-time cost regardless of the underlying generator.

While reranker-based defenses employ smaller models, their limited defensive effectiveness results in a less favorable security–efficiency trade-off. Overall, CanaryRAG achieves the strongest empirical protection while maintaining the lowest effective computational overhead among evaluated defenses.

Finally, in terms of end-to-end latency, reranker and summarization defenses introduce additional pre-generation stages that noticeably increase response time, while cascade decoding in RAGFort can nearly double latency and may induce generation artifacts. CanaryRAG is the only approach that operates fully online and concurrently, resulting in negligible additional latency compared to undefended inference.

\subsection{Summarization and Semantic Paraphrasing Adversaries}
A natural question concerns whether summarization or semantic paraphrasing constitutes a meaningful adaptive strategy against CanaryRAG. We consider that, under the standard knowledge base extraction threat model, such behaviors do not align with the attacker’s ultimate objective and therefore should be interpreted differently from extraction-oriented attacks.

In a black-box extraction setting, the adversary’s goal is to reconstruct the underlying knowledge base with high fidelity, typically at the granularity of documents or corpus. Summarization- or paraphrase-based outputs fundamentally conflict with this objective: they introduce substantial information loss and abstraction, resulting in large edit distances from the original chunks and yielding poor reconstruction quality. This limitation is intrinsic rather than defense-specific—no post-processing can reliably recover fine-grained structure, ordering, or technical detail from a summary-level disclosure.

Actually an extracted chunk is considered as a valid target chunk only if it satisfies \emph{both} of the following criteria\citep{di2024pirates}: (1) Lexical Similarity: the ROUGE-L score between the generated output and the original chunk exceeds $0.5$; and (2) Semantic Similarity: the cosine similarity between their embedding representations, computed using the same embedding model as the RAG agent’s retrieval pipeline, exceeds $0.85$. For bypass methods that obfuscate output, we hope that the output will also meet this standard after deobfuscation.

Moreover, LLM hallucination further weakens the attacker’s position. When only summarized or paraphrased content is available, \textbf{the adversary lacks a reliable mechanism to distinguish faithful compression of retrieved knowledge from hallucinated abstractions introduced by LLM}. This ambiguity is especially problematic when the extraction target is \textbf{literary works or technical documents, where summaries are insufficient} for reconstruction and cannot be validated \textbf{without} access to the original text. From an attacker’s perspective, such strategies are therefore better viewed as auxiliary techniques(In fact it has already been used in RAG-thief\cite{jiang2025feedback}) rather than primary extraction mechanisms. In practice, summarization and paraphrasing are more suitable for cold-start initialization in attacks such as Pirates or RAG-thief, where coarse semantic cues may help bootstrap subsequent iterative extraction. This interpretation is consistent with prior findings(in \Cref{sec:rq1}) that a strong initialization can improve downstream Chunk Recovery Rate (CRR). Importantly, our experimental protocol already incorporates such initialization, ensuring that CanaryRAG is evaluated under a realistically strong adversary.

Finally, we emphasize that these behaviors are not excluded from our evaluation. The adaptive attack results reported in \Cref{tab:rq4_adaptive_tpr}, specifically Adaptive (A2 + A3), explicitly include summarization- and paraphrase-style instructions alongside obfuscation strategies. The observed robustness therefore reflects CanaryRAG’s effectiveness even when such auxiliary adaptive behaviors are present.

\subsection{Future Work}
This work focuses on runtime integrity mechanism for detecting RAG knowledge base leakage. An important direction for future research is to investigate how CanaryRAG can be systematically combined with orthogonal defenses operating at retrieval time, decoding time, and post-generation filtering. Rather than serving as a replacement for existing safeguards, CanaryRAG is naturally complementary to these mechanisms. Such components can form a \textbf{defense-in-depth architecture} together. 

\section{The Philosophy of Canary Design in LLM}
CanaryRAG is motivated by an analogy to \emph{stack canaries} in software security~\citep{stack-canary}. In a correctly executing program, stack canaries are not accessed or modified by the program logic itself; their presence is orthogonal to functional execution and becomes observable only when control flow integrity is violated. This property enables reliable runtime detection of exploitation without interfering with normal program behavior. Our goal is to identify and operationalize an analogous property in LLMs.

In the context of retrieval-augmented generation, we seek signals that are \emph{present in the execution context but irrelevant to the task semantics}. Such signals should be effectively ignored during benign generation, yet become exposed when the model is induced to reproduce retrieved content. Through empirical exploration, we observe that inserting \textbf{high-entropy, random strings} into retrieved contexts exhibits precisely this behavior. These strings are constructed to avoid forming meaningful words or phrases and are not naturally present in the underlying RAG corpus.

Under normal user queries, the model almost never emits such high-entropy strings, as they carry no semantic relevance to the task. At the same time, their presence in the context does not materially affect generation quality or latency, a finding corroborated by our evaluations in RQ2 (\Cref{sec:rq2}) and RQ3 (\Cref{sec:rq3}). This indicates that high-entropy, context-irrelevant strings are effectively inert during benign execution.

This observation constitutes the core insight behind CanaryRAG: \emph{high-entropy, semantically irrelevant tokens embedded in retrieved context satisfy the defining properties of a canary in LLM-based systems}. They are ignored during normal operation, yet become difficult to suppress when extraction behavior forces faithful reproduction of the context. Consequently, canaries in CanaryRAG are not required to be secret, document-specific, or semantically meaningful. Instead, they function as lightweight integrity markers whose exposure is causally coupled with violations of the intended RAG execution pattern. This insight directly guides the subsequent design choices regarding canary placement and surface form.

\subsection{Canary Placement Strategies}
While the semantic irrelevance of canaries establishes their suitability as runtime integrity witnesses, their \emph{placement} within retrieved documents determines how reliably this witness is exposed under different extraction behaviors. 

A conservative placement strategy inserts canaries at \emph{inter-chunk boundaries}, i.e., between retrieved chunks rather than within their internal content. This approach preserves the internal semantics of each chunk and minimizes interference with downstream generation. However, this strategy may be less sensitive to fine-grained attacks that selectively extract partial content from within individual chunks. To address such cases,a more direct alternative is to place canaries \emph{within chunks}, thereby coupling canary exposure more closely with content-level reproduction behavior. Concretely, \emph{intra-chunk placement} can be further instantiated at different granularities, most notably at sentence boundaries, resulting in two representative strategies: sentence-begin and sentence-end insertion.

In practice, we empirically compare sentence-begin and sentence-end intra-chunk placement and observe no significant difference in detection effectiveness(as shown in \cref{tab:canary-placement-ablation}). However, sentence-end placement may allow partial content reproduction before canary exposure is triggered, as canaries appear only after sentence completion. For this reason, we adopt sentence-begin placement as a conservative default, ensuring earlier exposure under fine-grained extraction without materially affecting generation quality.

\begin{table}[t]
\centering
\small
\begin{tabular}{lcc}
\toprule
\textbf{Canary Placement} & \textbf{Non-adaptive} & \textbf{Adaptive} \\
\midrule
Inter-chunk (Boundary)      & 97.4\% & 88.4\% \\
Intra-chunk (Sentence-end) & 98.1\% & 93.0\% \\
Intra-chunk (Sentence-begin) & 98.2\% & 94.1\% \\
\bottomrule
\end{tabular}
\caption{Effect of canary placement strategies on detection effectiveness under non-adaptive and adaptive extraction attacks. All results are measured on the ChatDoctor agent with CanaryRAG enabled.}
\label{tab:canary-placement-ablation}
\end{table}

\section{Oracle Design Details}
\paragraph{Strategic fast-fail optimization} A key observation underlying our oracle design is that the oracle task does not need to faithfully reproduce the entire retrieval context to fulfill its role. Unlike the target path, which operates over the top-$k$ retrieved chunks to maximize task utility, the oracle path only needs to validate whether the model remains capable of exposing canaries. In practice, this property allows the oracle to operate on a \emph{single retrieved chunk} containing canaries, rather than the full retrieval set. Specifically, we construct the oracle context by selecting a random chunk from the top-k retrieved results and issuing an oracle prompt that explicitly induces faithful reproduction. If an adversary’s strategy successfully suppresses canary exposure in the target path, the same suppression also easily influenced to the canaries in oracle path.

\begin{figure}[t]
    \centering
    \includegraphics[width=7cm]{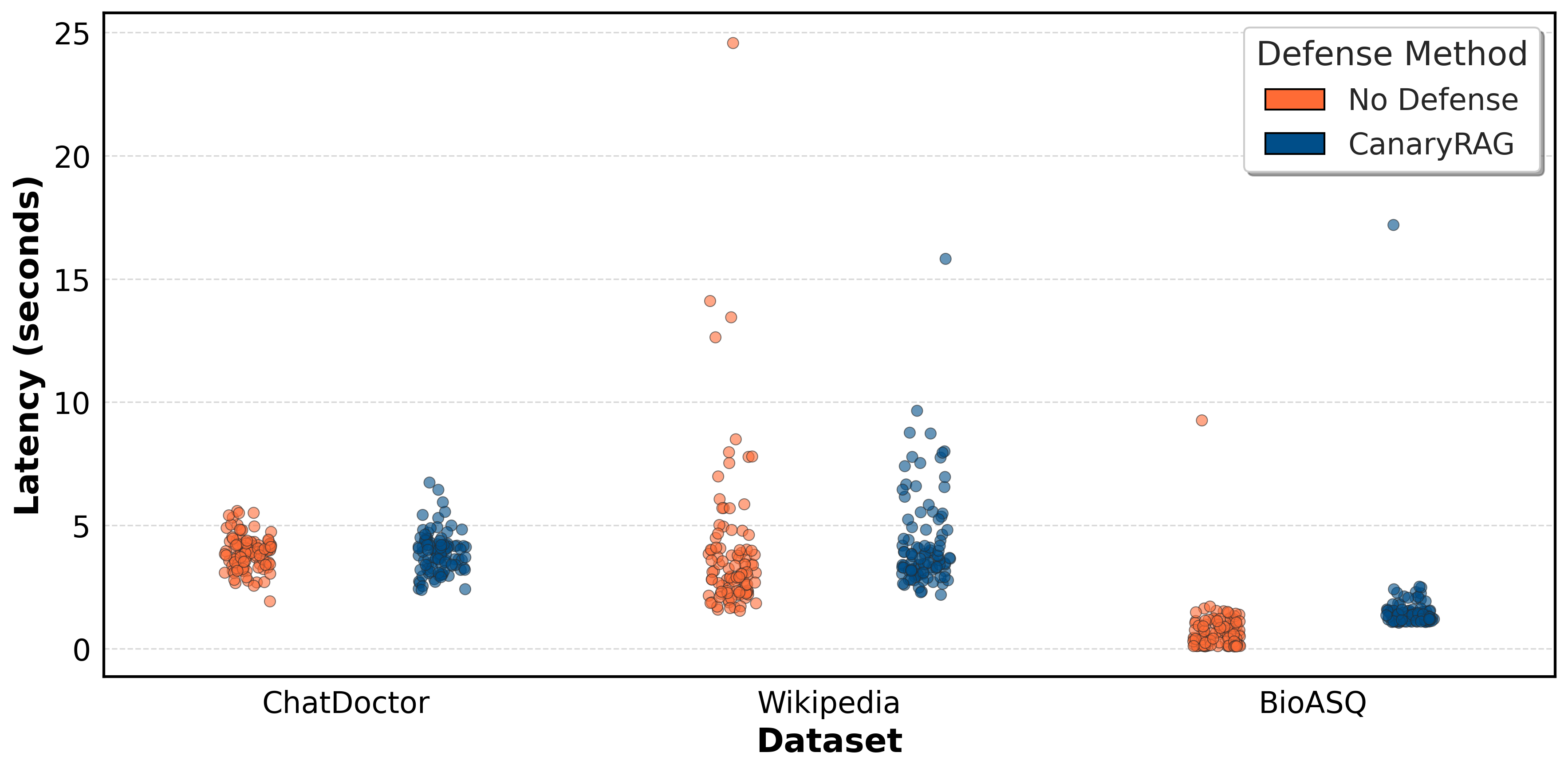}
    \caption{Per-query latency scatter plots across datasets with and without CanaryRAG.}
    \label{fig:latency_scatter}
\end{figure}

\section{Privacy-Utility Trade-off}
\label{sec:appendix_tradeoff}
\begin{table*}[h]
\centering
\small
\begin{tabular}{lcc}
\toprule
\textbf{Defense Method} & \textbf{Privacy (CRR$\downarrow$)} & \textbf{Utility (BERTScore$\uparrow$)} \\
\midrule
No Defense & 97.4 & 0.517 \\
Reranker   & 95.0 & 0.518 \\
Summarize  & 76.0 & 0.485 \\
RAGFort    & 50.2 & 0.493 \\
CanaryRAG  & \textbf{1.6} & 0.515 \\
\bottomrule
\end{tabular}
\caption{Privacy-utility trade-off of ChatDoctor Agent under worst-case attack setting.}
\label{tab:tradeoff}
\end{table*}

We evaluate the privacy-utility trade-off of different defense methods under a worst-case attack setting. We measure privacy using the Chunk Recovery Rate (CRR$\downarrow$), where lower values indicate stronger protection against knowledge base extraction. Utility is measured using BERTScore (BERTScore$\uparrow$), where higher values indicate better response quality. Table~\ref{tab:tradeoff} presents the results of all methods.

Among the baselines, Reranker provides negligible privacy improvement despite slightly improving utility, suggesting that retrieval-level adjustments alone are insufficient to mitigate extraction attacks. Summarize reduces leakage to some extent but introduces a noticeable drop in utility and additional computational overhead due to multi-step generation. RAGFort achieves moderate privacy gains but at the cost of reduced utility, indicating a less favorable trade-off. CanaryRAG maintains near-optimal utility, comparable to No Defense and significantly better than Summarize and RAGFort. This demonstrates that CanaryRAG achieves Pareto optimality in the privacy-utility trade-off across all baselines.
\end{document}